\documentclass[iop]{emulateapj-rtx4}
\usepackage{natbib}
\usepackage{color}
\usepackage{graphicx}
\newcommand{\Ka}{K$\alpha$}
\newcommand{\Kb}{K$\beta$}

\newcommand{\Msun}{$M_{\odot}$}

\shorttitle{Electron Capture Products in Type Ia SNR 3C\,397}
\shortauthors{Yamaguchi et al.}

\begin{document}

\title{A Chandrasekhar Mass Progenitor for the Type Ia Supernova Remnant 3C\,397 \\
	from The Enhanced Abundances of Nickel and Manganese}

\author{
Hiroya Yamaguchi\altaffilmark{1,2,3},
Carles Badenes\altaffilmark{4}, 
Adam R.\ Foster\altaffilmark{3}, 
Eduardo Bravo\altaffilmark{5}, 
Brian J.\ Williams\altaffilmark{1},\\
Keiichi Maeda\altaffilmark{6,7},
Masayoshi Nobukawa\altaffilmark{8},
Kristoffer A.\ Eriksen\altaffilmark{9}, 
Nancy S.\ Brickhouse\altaffilmark{3},\\
Robert Petre\altaffilmark{1},
Katsuji Koyama\altaffilmark{8,10}
}
\email{hiroya.yamaguchi@nasa.gov}

\altaffiltext{1}{NASA Goddard Space Flight Center, Code 662, Greenbelt, MD 20771, USA}
\altaffiltext{2}{Department of Astronomy, University of Maryland, College Park, MD 20742, USA}
\altaffiltext{3}{Harvard-Smithsonian Center for Astrophysics, 60 Garden Street, 
	Cambridge, MA 02138, USA}
\altaffiltext{4}{Department of Physics and Astronomy and Pittsburgh Particle Physics, 
	Astrophysics and Cosmology Center (PITT PACC), University of Pittsburgh, 
	3941 O'Hara St, Pittsburgh, PA 15260, USA}
\altaffiltext{5}{E.T.S.\ Arquitectura del Vall{\`e}s, Universitat Polit{\`e}cnica de Catalunya, 
  Carrer Pere Serra 1-15, 08173 Sant Cugat del Vall{\`e}s, Spain}
\altaffiltext{6}{Department of Astronomy, Kyoto University, Kitashirakawa-oiwake-cho, Sakyo-ku, 
	Kyoto 606-8502, Japan}
\altaffiltext{7}{Kavli Institute for the Physics and Mathematics of the Universe (WPI), 
	University of Tokyo, 5-1-5 Kashiwanoha, Kashiwa, Chiba 277-8583, Japan}
\altaffiltext{8}{Department of Physics, Kyoto University, Kitashirakawa-oiwake-cho, Sakyo-ku, 
	Kyoto 606-8502, Japan}
\altaffiltext{9}{Theoretical Design Division, Los Alamos National Laboratory, PO Box 1663, 
	Los Alamos, NM 87545, USA}
\altaffiltext{10}{Department of Earth and Space Science, Osaka University, 1-1 Machikaneyama, 
	Toyonaka, Osaka 560-0043, Japan}

\begin{abstract}
  Despite decades of intense efforts, many fundamental aspects of Type Ia supernova (SNe Ia) remain elusive. One of the major open
  questions is whether the mass of the exploding white dwarf (WD) is close to the Chandrasekhar limit.  Here we report the detection of
  strong K-shell emission from stable Fe-peak elements in the {\it Suzaku} X-ray spectrum of the Type Ia supernova remnant (SNR)
  3C\,397.  The high Ni/Fe and Mn/Fe mass ratios (0.11--0.24 and 0.018--0.033, respectively) in the hot plasma component that dominates  
  the K-shell emission lines indicate a degree of neutronization
  in the SN ejecta which can only be achieved by electron captures in the dense cores of exploding WDs with a near-Chandrasekhar
  mass.  This suggests a single-degenerate origin for 3C\,397, since Chandrasekhar mass progenitors are expected naturally if the
  WD accretes mass slowly from a companion. Together with other results supporting the double-degenerate scenario, our work adds
  to the mounting evidence that both progenitor channels make a significant contribution to the SN Ia rate in star-forming
  galaxies.
\end{abstract}


\section{Introduction}

Type Ia supernovae (SNe Ia) are widely believed to result from the thermonuclear explosion of a carbon-oxygen white dwarf (WD)
that is destabilized by mass transfer in a binary system \citep[e.g.,][]{Maoz14}.  Even though their use as distance indicators in
cosmology has sparked considerable interest in SNe Ia \citep[e.g.,][]{Riess98,Perlmutter99}, many fundamental aspects of these
explosions remain obscure.  Two major channels are thought to lead to SN Ia explosions: the single degenerate (SD) scenario
\citep{Whelan73} where a WD accretes matter from a non-degenerate companion and explodes when its mass grows close to the
Chandrasekhar limit ($M_{\rm Ch} \approx 1.4$\,\Msun), and the double degenerate (DD) scenario \citep{Webbink84} where the
explosion is triggered by the dynamical merger of two WDs.  Unfortunately, the remarkable uniformity in SN Ia light curves and
spectra makes it difficult to infer the properties of the progenitors from the explosions themselves.  Because of this, most
observational efforts to distinguish between progenitor scenarios for individual SNe Ia have focused on searches for
circumstellar material \citep{Patat07,Sternberg11,Badenes07,Williams11,Williams14} or pre-existing/surviving stellar companions
\citep{Li11,Gonzalez12,Schaefer12}.

From the point of view of the SN nucleosynthesis, the main difference between SD and DD systems is the central density of the WD
at the onset of the thermonuclear runaway \citep{Pakmor12,Seitenzahl13b}.  In SD progenitors, the exploding WD should always be
close to $M_{\rm Ch}$, and have a dense ($\rho \gtrsim 2 \times 10^{8}$\,g\,cm$^{-3}$) core where electron captures can take place
efficiently, leading to significant production of neutronized species like $^{58}$Ni and $^{55}$Mn
\citep{Iwamoto99,Seitenzahl13a}.  In contrast, the DD explosion scenarios that can best reproduce the observed properties of SNe
Ia require lower masses and central densities for the primary WD, and therefore predict lower yields of these species
\citep{Pakmor12,Kerkwijk13}.  For this reason, the yield of $^{58}$Ni, $^{55}$Mn, and other neutronized species has been identified as a powerful
diagnostic for SN Ia progenitors \citep{Maeda10b,Seitenzahl15}.  However, a combination of long half-lives and complex ionization
issues at play in nebular spectra \citep{Dessart14} makes it extremely difficult to quantify these yields for individual objects during the
optical SN phase \citep{Mazzali07,Gerardy07,Maeda10a}.

X-ray observations of evolved supernova remnants (SNRs) offer an excellent opportunity to make robust measurements of 
the yields of neutronized species, as the innermost ejecta must have been thermalized by the reverse shock in the SNRs, 
and will therefore be visible in the X-ray spectrum. 3C\,397 (G41.1--0.3) is an ideal target in this sense, since it is dynamically 
more evolved than other well-known Type Ia SNRs (i.e., Kepler, Tycho, SN\,1006).  Although 3C\,397 has sometimes been 
classified as a core-collapse SNR due to its proximity to molecular clouds \citep{Safi00}, most studies, including our own 
systematic analysis of Fe K-shell emission in young and middle-aged SNRs, agree on a Type Ia origin 
\citep[e.g.,][]{Chen99,Yang13,Yamaguchi14b}. Here we present strong evidence for the presence of 
electron capture products in the X-ray spectrum of 3C\,397, after proving that the SNR is indeed evolved.
In our analysis, we assume the distance to the SNR to be 10\,kpc \citep[][and references therein]{Safi00}, 
but our main results and conclusions are not affected by the uncertainty in the distance estimate.

The errors quoted in the text and table, and the error bars in the figures are 
at the 1$\sigma$ confidence level, unless otherwise stated.

\section{Observational Results}

\begin{figure}[t]
  \begin{center}
	\vspace{1mm}
	\includegraphics[width=7.2cm]{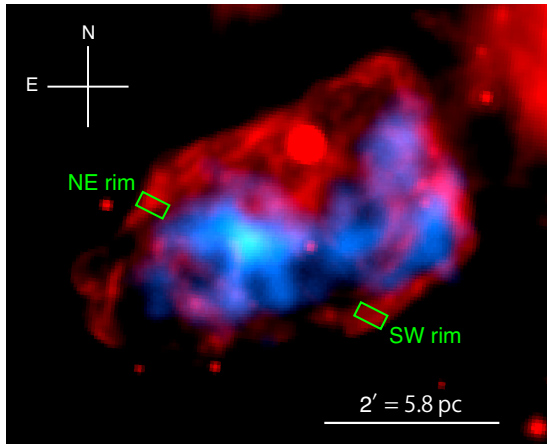}	
	\vspace{1mm}
\caption{
  Two-color image of SNR 3C\,397.
  Red and blue represent 24\,$\mu$m IR ({\it Spitzer}) and 5--9\,keV X-rays 
  ({\it Suzaku}), respectively.  The green rectangles indicate the locations 
  where the IR spectra (Fig.\,\ref{fig:spec}a) are extracted. 
  \label{fig:image}}
  \end{center}
\end{figure}

We analyzed archival  {\it Spitzer} infrared (IR) and {\it Suzaku} X-ray data from 3C\,397. 
The IR observations were performed using the Multiband Imaging Photometer (MIPS) and 
the Infrared Spectrograph (IRS) during April 2005. The X-ray observation was performed in October 2010 
with a total effective exposure of 69\,ks for the X-ray Imaging Spectrometer (XIS).
We followed the standard procedures for data reduction. 
Figure\,\ref{fig:image} shows a composite image of
3C\,397 in the 24\,$\mu$m IR (red) and 5--9\,keV X-rays (blue).

\begin{figure}[t]
  \begin{center}
	\vspace{1mm}
	\includegraphics[width=7.4cm]{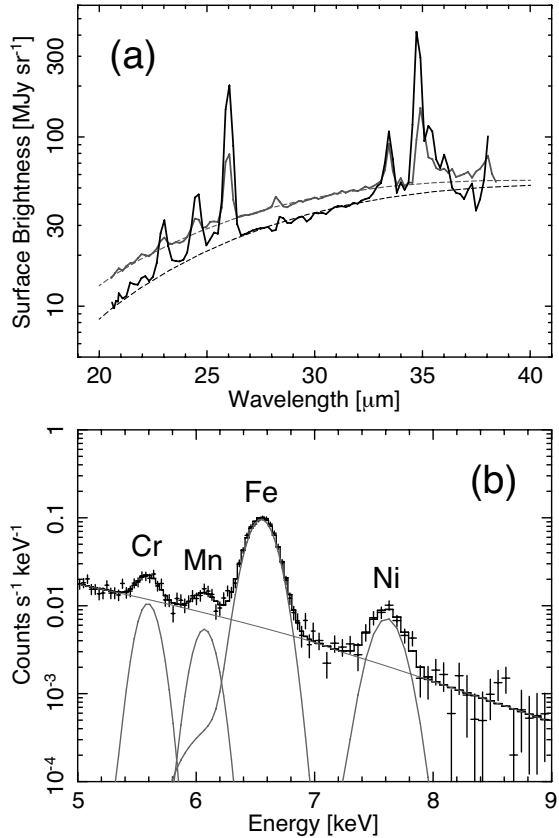}	
	\vspace{1mm}
\caption{
  (a) {\it Spitzer} IRS spectra used to measure the ambient density.
  Black and gray data are taken from the NE and SW rims indicated 
  in Figure\,\ref{fig:image}, respectively.  
  The continuum components are fitted with a dust emission model (dashed curves). \
  (b) Spatially integrated {\it Suzaku} XIS spectrum of 3C\,397. 
  The model (gray lines) consists of a bremsstrahlung continuum with 
  an electron temperature of $log_{10}(T_e\,{\rm [K]}) = 7.39 \pm 0.04$ 
  and four Gaussians with the parameters listed in Table\,\ref{tab:best}.
  \label{fig:spec}}
  \end{center}
\end{figure}

\begin{table*}[t]
\begin{center}
\caption{The best-fit spectral parameters, theoretically-predicted line emissivities 
and their ratios, and derived mass ratios for the Fe-peak elements observed in 
the hard (5--9\,keV) X-ray spectrum of 3C\,397.
  \label{tab:best}}
  \begin{scriptsize}
  \begin{tabular}{lcccccccc}
\hline \hline
~ & Centroid & FWHM & Flux & & 
 Emissivity & $\varepsilon / \varepsilon _{\rm Fe}$ & $M / M_{\rm Fe}$ \\
~ & (eV) & (eV) & (photons\,cm$^{-2}$\,s$^{-1}$)  & & ($10^{-13}$\,photon\,cm$^3$\,s$^{-1}$) & ~ & ~ \\
\cline{1-4} \cline{6-8}
Cr \Ka\  & $5596_{-11}^{+12}\,(\pm 6)$ & $104_{-64}^{+45}$ & $1.05_{-0.14}^{+0.15} \times 10^{-5}$ 
& & $3.3_{-1.0}^{+1.1}$ &  $2.6 \pm 0.2$ & $0.027_{-0.006}^{+0.007}$ \\
Mn \Ka\ & $6073_{-20}^{+19}\,(\pm 6)$ & $104$ & $5.75_{-1.10}^{+1.17} \times 10^{-6}$ 
& & $2.2 \pm 0.7$ &  $1.7 \pm 0.1$ &  $0.025_{-0.007}^{+0.008}$ \\
Fe \Ka\ & $6556_{-2}^{+3}\,(\pm 7)$ & $181 \pm 6$ & $1.38 \pm 0.03  \times 10^{-4}$ 
& & $1.3_{-0.4}^{+0.5}$ & 1 & 1 \\
Ni \Ka\ (+ Fe \Kb) & $7616 \pm 13\,(\pm 8)$ & $197_{-47}^{+46}$ & $1.61_{-0.17}^{+0.18} \times 10^{-5}$ 
& & --- & --- & --- \\
Ni \Ka\ & --- & --- & $1.09_{-0.30}^{+0.34} \times 10^{-5}$ 
& & $0.64_{-0.22}^{+0.26}$ &  $0.49 \pm 0.03$ & $0.17_{-0.05}^{+0.07}$ \\
\hline
\end{tabular}
\end{scriptsize}
\tablecomments{
The uncertainties in parenthesis are the systematic component 
\citep[0.1\% of the mean energy;][]{Ozawa09a}. 
The width (FWHM) of the Mn \Ka\ line was linked to that of Cr \Ka. 
The line emissivity $\varepsilon$ is defined as 
$F = \varepsilon \, n_e n_{\rm ion} V / (4 \pi D^2)$, where $F$, $n_{\rm ion}$, $V$, and $D$, 
are the line flux, ion number density, emitting volume, and distance to the SNR, respectively. 
The Ni \Ka\ emission observed in the X-ray spectrum (Fig.\,\ref{fig:spec}b) is in fact contributed 
by Fe \Kb\ lines as well, with their centroids overlapping with each other (see text for details). 
The Ni \Ka\ flux and Ni/Fe mass ratio given in the last row are obtained after subtracting 
the Fe \Kb\ contribution. 
}
\end{center}
\end{table*}

\subsection{IR Spectroscopy}

In young SNRs, mid-IR continuum emission arises from dust grains in the interstellar medium (ISM) that are heated in the
post-shock gas by collisions with energetic ions and electrons. Therefore, IR spectra can constrain the density and mass of the
swept-up ambient medium \citep[e.g.,][]{Borkowski06}.  Figure\,\ref{fig:spec}a shows the background-subtracted {\it Spitzer} IRS
spectra from the two regions indicated in Figure\,\ref{fig:image}.  Using the dust emission models and analysis procedures
described in \cite{Borkowski06} and \cite{Williams12}, we fitted the 21--33\,$\mu$m continuum and derived post-shock proton number
densities of $4.6 \pm 0.4$\,cm$^{-3}$ and $8.5 \pm 0.8$\,cm$^{-3}$ for the NE and SW rims, respectively.  The average pre-shock density is therefore $\rho_0 =
\mu _{\rm H} \times (4.6 + 8.5) / 2 \times (1/4)$ $\approx$ $3.8 \times 10^{-24}$\,g\,cm$^{-3}$, where $\mu _{\rm H}$ = $1.4 \times 1.67
\times 10^{-24}$\,g is the mean mass per hydrogen nucleus for solar abundances \citep{Asplund09}.

The spatially-integrated IR flux from the SNR is $\sim$\,20\,Jy, which requires a swept-up dust mass of
$\sim$\,0.2\,$M_{\odot}\,d_{10}^2$. This estimate accounts for the fact that the mid-IR observations usually capture only
$\sim$20\% of the flux in this band, because the rest of the dust is colder and emits at longer wavelengths
\citep{Borkowski06,Williams14}.  Taking the standard dust-to-gas ratio in the Milky Way \citep{Weingartner01}, the estimated dust
mass leads to a swept-up gas mass of $\sim$\,25\,$M_{\odot}\,d_{10}^2$.  The SNR shape can be well approximated by a $2.\!'4
\times 1.\!'3 \times 1.\!'3$ ellipsoid, corresponding to a volume of $1.4 \times 10^{58}\,d_{10}^{3}$\,cm$^3$.  The average
ambient density that the SNR blast wave has experienced is therefore estimated to be $3.6 \times
10^{-24}\,d_{10}^{-1}$\,g\,cm$^{-3}$, consistent with (and independent from) the measurement from the IR spectra at the nominal
distance to the SNR.  Because the total swept-up mass ($\sim$\,25\,$M_{\odot}$) is much larger than the typical SN Ia ejecta mass,
we conclude that the reverse shock must have thermalized the innermost ejecta \citep[e.g.,][]{Truelove99}.

\subsection{X-Ray Spectroscopy}

Given the ISM origin of the IR emission, the clear anticorrelation between the IR and X-ray morphologies (Fig.\,\ref{fig:image})
suggests that the hard X-rays predominantly originate from the ejecta. The spectrum of the {\it Suzaku} XIS in the 5--9\,keV band
extracted from the entire SNR is shown in Fig.\,\ref{fig:spec}b.  The data from the two active front-illuminated CCDs (XIS0 and 3)
were merged after background subtraction. The K$\alpha$ atomic lines from four Fe-peak elements (Cr, Mn, Fe, and Ni) are clearly
resolved and detected at high significance. We measured their centroids and fluxes using Gaussian models, and obtained the results
given in Table\,\ref{tab:best}.  
Since the observed spectrum is dominated by metal-rich ejecta, we modeled the continuum with 
a bremsstrahlung representing emission from collisions between hot free electrons and highly-ionized 
heavy elements by using the Gaunt factor calculation method described in \cite{Kellogg75}. 
This model constrains the electron temperature to $log_{10}(T_e\,{\rm [K]}) = 7.39 \pm 0.04$.  
If a power law model is adopted instead, an unreasonably steep photon index ($\Gamma
\gtrsim 4.0$) is obtained, ruling out a nonthermal origin for the continuum. 
In fact, we found no evidence for a synchrotron X-ray filament, such as observed in young SNRs (e.g., Tycho, SN\,1006), 
in the higher-resolution {\it Chandra} image of 3C\,397 at the line-free energies of 4.1--5.3\,keV. 
During the spectral analysis, we kept the absorption column density fixed at $N_{\rm H} = 
3 \times 10^{22}$\,cm$^{-2}$ \citep{Safi05} with standard Galactic abundances \citep{Wilms00}.  
We repeated the analysis using different background spectra, 
but found no significant change in the measured values listed in Table\,\ref{tab:best}.

\begin{figure*}[t]
  \begin{center}
	\includegraphics[width=17.2cm]{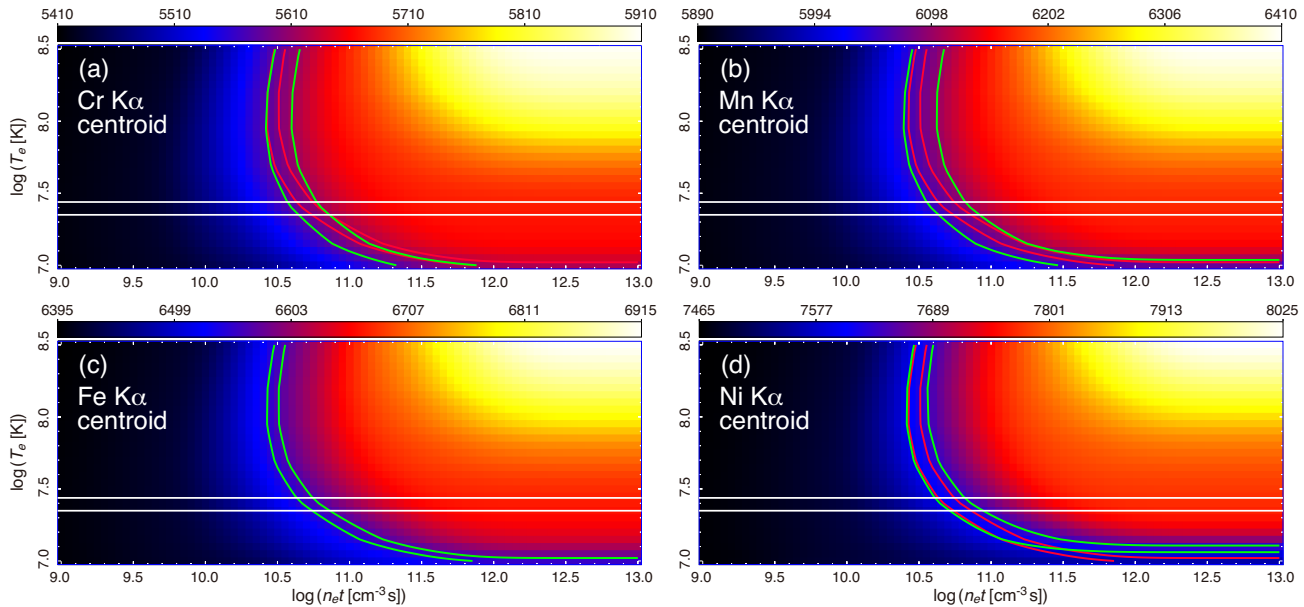}	
	\vspace{1mm}
\caption{
  Theoretical predictions for the centroid energies of K$\alpha$ emission from 
  (a) Cr K$\alpha$, (b) Mn K$\alpha$, (c) Fe K$\alpha$, and (d) Ni K$\alpha$, 
  as a function of the ionization timescale ($n_e t$: horizontal axis) 
  and electron temperature ($T_e$: vertical axis). 
  The regions constrained by the observed centroid energies are indicated by 
  the green curves. The red contours in panels (a), (b), and (d) are identical to 
  the green contours in panel (c).  The horizontal white lines indicate the electron
  temperature constrained by the shape of the bremsstrahlung continuum.  The overlap of the contours indicates that the emission from all four lines originates in a single plasma component with an ionization timescale of $log_{10}(n_e t\,[{\rm cm}^{-3}\,{\rm s}]) = 10.73 \pm 0.10$.
Electronic data of the centroid energies and emissivities we used are available at $<$http://asd.gsfc.nasa.gov/Hiroya.Yamaguchi/AtomicData/$>$.
  \label{fig:atomic}}
  \end{center}
\end{figure*}

In the non-equilibrium ionization (NEI) conditions commonly found in SNRs, line centroids and emissivities are determined by the
electron temperature ($T_e$) and the ionization timescale ($n_{e}t$ -- the product of the electron number density and plasma age,
which is the time elapsed since the gas was shock-heated).  To determine the plasma properties and elemental mass ratios from the
observed X-ray spectrum, we computed new atomic data based on the updated {\it AtomDB} database\footnote{http://www.atomdb.org}
(A.\ R.\ Foster et al., in preparation).  Figure\,\ref{fig:atomic} shows theoretical centroid energies for the Fe-peak elements 
detected in the hard X-ray spectrum as a function of ionization timescale (horizontal axis) and electron temperature (vertical axis). 
The regions constrained by the observed spectrum are indicated in the plots.  We find that the plasma conditions for the four 
elements overlap with one another, indicating that they originate in a single plasma component with an ionization timescale 
of $log_{10}(n_e t\,[{\rm cm}^{-3}\,{\rm s}]) = 10.73 \pm 0.10$. 
The K$\alpha$ line emissivities and derived mass ratios (relative to Fe) for the constrained plasma state 
are given in Table\,1. None of the values listed there depends on the distance to the SNR. 
For the conditions found in 3C\,397, the centroid of Fe K$\beta$ emission is predicted to be 
$7601_{-25}^{+28}$\,eV, overlapping with the measured range of the Ni K$\alpha$ centroid 
\citep[see][for Fe \Kb\ atomic data]{Yamaguchi14a}. 
This indicates that Fe K$\beta$ emission contaminates the Ni K$\alpha$ flux. Therefore, we calculated the Fe K$\beta$/K$\alpha$ emissivity ratio, 
obtaining $\varepsilon _{{\rm Fe(K}\beta{\rm )}}/\varepsilon _{{\rm Fe(K}\alpha{\rm)}}$ = $3.8_{-0.5}^{+0.4}$\%. 
The Ni/Fe mass ratio given in Table\,1 was derived taking this contribution from the Fe K$\beta$ lines 
into consideration.

\section{Discussion}

The classification of SNRs as Type Ia or core collapse can sometimes be controversial, 
but the observational evidence for 3C\,397 strongly favors a SN Ia origin 
\citep[e.g.,][]{Chen99,Yang13,Yamaguchi14b}. In particular, the overall abundance pattern 
revealed by {\it Chandra} data \citep[e.g., Fe/Mg $\gtrsim$ 10\,solar, Fe/Si $\gtrsim$ 3\,solar;][]{Safi05} is consistent 
with typical SN Ia yields \citep[e.g.,][]{Iwamoto99}. Although \cite{Safi00} did propose 
a core-collapse origin for this SNR, their classification was based on the high ambient 
density ($\sim 1 \times 10^{-22}$\,g\,cm$^{-3}$) estimated from the soft X-ray spectrum 
in {\it ASCA} data. This technique can yield highly uncertain results in heavily absorbed 
objects like 3C\,397. In fact, our IR observations indicate that the ambient density is 
much lower (\S2.1), more in line with other known Type Ia objects \citep{Badenes07}. 
The IR results are also consistent with our recent systematic study of Fe K emission in SNRs, 
where 3C\,397 is placed squarely in the SN Ia region with an ambient density less 
than $5 \times 10^{-24}$\,g\,cm$^{-3}$\citep{Yamaguchi14b}.

\begin{figure}[t]
  \begin{center}
	\vspace{1mm}
	\includegraphics[width=7.2cm]{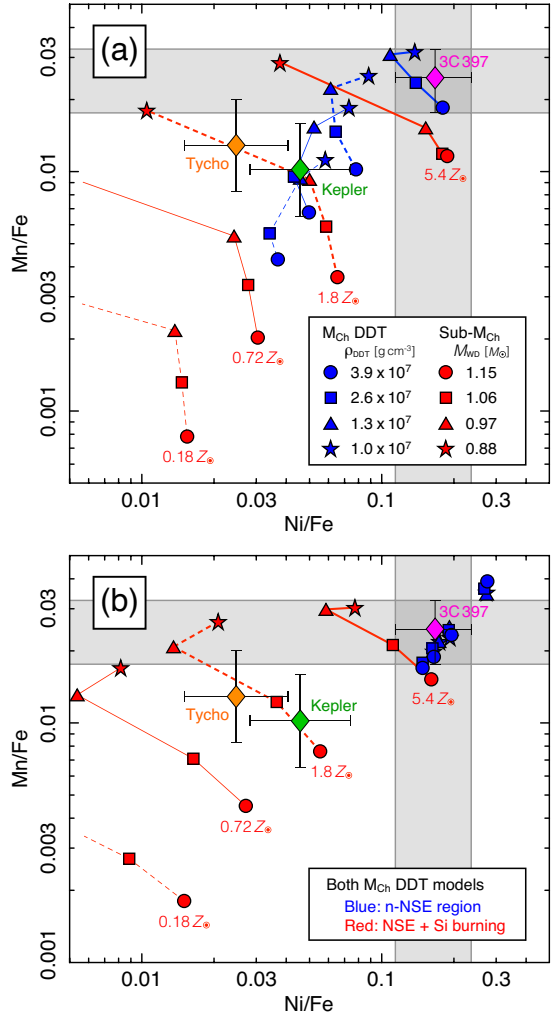}	
	\vspace{1mm}
\caption{
    (a) Ni/Fe and Mn/Fe mass ratios in SN Ia models, compared with the observed values 
    in 3C\,397 (magenta diamond and gray regions).
    Blue symbols represent $M_{\rm Ch}$ delayed-detonation models with different values of 
    $\rho_{\rm DDT}$ as indicated in the legend. 
    Red symbols represent sub-$M_{\rm Ch}$ detonation models with WD masses given in the legend. 
    Models with the same progenitor metallicity are connected by lines: 
    5.4\,$Z_\odot$ (thick solid), 1.8\,$Z_\odot$ (thick dashed), 
    0.72\,$Z_\odot$ (thin solid), and 0.18\,$Z_\odot$ (thin dashed), 
    The values for Kepler and Tycho indicated by the green and orange diamonds
    are calculated using the line fluxes from \cite{Park13} and \cite{Yamaguchi14a} 
    with our updated atomic data. \ 
    (b) Same as panel (a), but the values predicted for the innermost 0.2\,\Msun\ 
    that is dominated by the n-NSE regime (blue)  
    and the other regions (red) from the $M_{\rm Ch}$ models are separately shown. 
    The observed mass ratios for 3C\,397 can be well explained by the nucleosynthesis 
    that occurs in the n-NSE regions of standard metallicity models.
    \label{fig:abundance}
   }
  \end{center}
	\vspace{-1.5mm}
\end{figure}

In SNe Ia, Fe-peak nuclei are synthesized in three different burning regimes: incomplete Si burning, nuclear statistical
equilibrium (NSE), and neutron-rich NSE (n-NSE). In the incomplete Si burning and NSE regimes, the yield of neutronized species
(mainly stable $^{58}$Ni and $^{55}$Mn after the radioactive decays of $^{55}$Co$\rightarrow$$^{55}$Fe$\rightarrow$$^{55}$Mn) is
controlled by the pre-explosion neutron excess carried by $^{22}$Ne in the WD, which in turn is set mainly by the metallicity of the WD progenitor
\citep{Timmes03,Badenes08b,Bravo13,Park13}.  Only in the case of near-$M_{\rm Ch}$ WDs, the innermost $\sim$0.2$\,M_{\odot}$,
is consumed in the n-NSE regime \citep[e.g.,][]{Iwamoto99}, where density-driven electron captures generate a neutron excess
independently of progenitor metallicity.  The Ni/Fe mass ratio found in 3C\,397 (0.11--0.24, which is independent from 
the distance to the SNR) is, to our knowledge,
the highest reported in any SN Ia observation, and it can be produced only in the n-NSE regime (and thus only in the near-$M_{\rm Ch}$ WD) at 
near-solar metallicities (Fig.\,2b of Park et al.\ 2013). 

To explore the relationship between the progenitor properties and the yields of the Fe-peak elements in more detail, we calculated
a grid of SN Ia explosion models with a variety of progenitor WD masses ($M_{\rm WD}$ = 0.88\,$M_{\odot}$, 0.97\,$M_{\odot}$,
1.06\,$M_{\odot}$, 1.15\,$M_{\odot}$, and $M_{\rm Ch}$ $\approx$ 1.37\,$M_{\odot}$) and metallicities reasonable 
for Milky Way stars \citep{Rocha00} ($Z$ = 0.18\,$Z_{\odot}$,
0.72\,$Z_{\odot}$, 1.8\,$Z_{\odot}$, and 5.4\,$Z_{\odot}$, where the updated value of \cite{Asplund09} is used for $Z_{\odot}$).
The models were calculated with a version of the one-dimensional code described in \cite{Bravo12}, updated to include a more
accurate treatment of the coupling between hydrodynamics and nuclear reactions (E.\ Bravo et al., in preparation).  For the
$M_{\rm Ch}$ cases, delayed-detonation models \citep{Khokhlov91} with various deflagration-to-detonation transition densities
($\rho_\mathrm{DDT}$) were used.  The sub-$M_{\rm Ch}$ explosions were initiated as detonations at the WD center \citep{Sim10},
which is a good approximation for a SN Ia explosion initiated by the violent merging of binary WDs \citep{Pakmor12}.  The mass of
$^{56}$Ni synthesized in the different models is between 0.17\,M$_\odot$ and 0.95\,M$_\odot$, in agreement with the range found in
normal SN Ia \citep{Scalzo14}.  As expected, the highest level of neutronization in the $M_{\rm Ch}$ models is achieved in the
n-NSE regime. We note that this result is mainly driven by the core density of the WD, and hence one-dimensional SN Ia models
should capture the fundamental trends in the synthesis of neutronized species.

In Figure\,\ref{fig:abundance}a, the Ni/Fe and Mn/Fe mass ratios predicted by these 
explosion models are compared with the observed values.  All the sub-$M_{\rm Ch}$ models, 
regardless of progenitor mass or metallicity, fail to reproduce the high levels of neutronization 
found in 3C\,397.  We find that the $M_{\rm Ch}$ models can match the observed mass ratios 
of both Ni/Fe and Mn/Fe, but they also require relatively high metallicities.  
A possible interpretation for this fact is that the hot plasma component responsible for 
the observed K-shell emission is dominated by the n-NSE products, whereas the majority 
of the NSE and Si-burning products composes a lower-temperature component 
that is visible in the soft X-ray band (i.e., L-shell emission). 
An alternative is that the emission from the n-NSE region is enhanced due to 
density inhomogeneities in the ejecta. Otherwise, the metallicity should be indeed high. 
Figure\,\ref{fig:abundance}b shows the mass ratios for the same $M_{\rm Ch}$ models 
where the values predicted for the n-NSE region (the innermost 0.2$\,M_{\odot}$ 
in our one-dimensional models) and the other regions are split. The mass ratios determined 
from the K-shell spectra can be well explained by the n-NSE components even with 
the relatively low metallicities, either alone or partially mixed with the rest of NSE matter. 
Since the n-NSE regime is not expected in sub-$M_{\rm Ch}$ WDs, we can conclude that 
the progenitor of 3C\,397 must have had a mass very close to $M_{\rm Ch}$.  
A $M_{\rm Ch}$ progenitor is naturally explained by the evolution of a WD slowly accreting mass 
from a non-degenerate companion \citep[e.g.,][]{Hachisu96}. Therefore, our results strongly 
suggest the SD scenario as the origin of this particular SN Ia.  In principle, $M_{\rm Ch}$ or even 
super-$M_{\rm Ch}$ WDs could arise in the DD scenario \citep{Howell06}, but the properties of 
the galactic population of WD binaries make this a remote possibility at best \citep{Badenes12}.

The large amount of neutronized material revealed by the X-ray spectrum of 3C\,397 might seem 
exceptional in comparison to other SN Ia remnants like Kepler \citep{Park13} or Tycho \citep{Yamaguchi14a} 
(see Figure\,\ref{fig:abundance}), but it is important to emphasize that 3C\,397 is the only evolved 
Type Ia SNR that has been observed to such depth by {\it Suzaku}.  
This implies that SD progenitors might be common in the Milky Way, which is also suggested from  
the presence of circumstellar material confirmed in some young SN Ia remnants 
\citep[e.g.,][]{Williams11,Williams12,Williams14}.
Since the evidence in favor of DD progenitors is strong for other SNRs \citep{Gonzalez12,Schaefer12}, 
it appears that both progenitor channels must contribute significantly to the SN Ia rate in star-forming galaxies.

\section{Conclusions}

We have shown that the SN Ia progenitor of 3C\,397 likely contained a WD with a mass very close to $M_{\rm Ch}$. This result is anchored
by the strong K-shell emission from Ni and Mn in this SNR, and is robust to the details of the data analysis and the SN nucleosynthesis 
calculations used to interpret the data.  Other work has claimed evidence for $M_{\rm Ch}$ SN Ia progenitors by modeling Galactic
chemical evolution \citep{Seitenzahl13b}, or by applying phenomenological radiative transfer relations to large samples of SN Ia
lightcurves \citep{Scalzo14}, but these studies make strong assumptions about complex and highly uncertain processes. 
The analysis of the hard X-ray spectrum of 3C\,397 presented here might be the cleanest, most robust determination of 
the mass of a single SN Ia progenitor to date, and strongly suggests an SD progenitor for this particular remnant. 
Future deep observations with higher angular/spectral resolution including the soft X-ray band will allow us to investigate 
the spatial distribution of the elemental mass ratios and plasma properties (i.e., electron temperature and density). 
This will help understand why the NSE and Si-burning products are little visible in the hard X-ray band (see \S3), 
and constrain the detailed explosion mechanism of $M_{\rm Ch}$ SNe Ia as well as the dynamical evolution of their remnants.

\acknowledgments
We thank Drs.\ Ken'ichi Nomoto, Samar Safi-Harb, Randall K.\ Smith, 
and Michael C.\ Witthoeft for helpful discussion and suggestions. 
E.B.\ is supported by Spanish MINECO grant AYA2013-40545. 
Japanese authors acknowledge financial support by JSPS Grant-in-Aid for Scientific 
Research numbers 23740141/26800100 (K.M.), 24740123 (M.N.), and 
23000004/24540229 (K.K). 

\bigskip


\end{document}